# Improved electronic measurement of the Boltzmann constant by Johnson noise Thermometry

Jifeng Qu[1], Samuel P Benz[2], Alessio Pollarolo[2], Horst Rogalla[2],
Weston L Tew[3], Rod White[4], and Kunli Zhou[1,5]

[1] *National Institute of Metrology (NIM), Beijing 100029, People's Republic of China*
[2] *National Institute of Standards and Technology (NIST), 325 Broadway, Boulder, CO 80305-3328, USA*[1]
[3] *National Institute of Standards and Technology (NIST), 100 Bureau Drive, Gaithersburg, MD 20899, USA*
[4] *Measurement Standards Laboratory of New Zealand, Lower Hutt, New Zealand*
[5] *Tsinghua University, Beijing 100084, People's Republic of China*

**Abstract**

The unit of thermodynamic temperature, the kelvin, will be redefined in 2018 by fixing the value of the Boltzmann constant, $k$. The present CODATA recommended value of $k$ is determined predominantly by acoustic gas-thermometry results. To provide a value of $k$ based on different physical principles, purely electronic measurements of $k$ were performed by using a Johnson noise thermometer to compare the thermal noise power of a 200 Ω sensing resistor immersed in a triple-point-of-water cell to the noise power of a quantum-accurate pseudo-random noise waveform of nominally equal noise power. Measurements integrated over a bandwidth of 550 kHz and a total integration time of 33 days gave a measured value of $k = 1.3806514(48) \times 10^{-23}$ J/K, for which the relative standard uncertainty is $3.5 \times 10^{-6}$ and the relative offset from the CODATA 2010 value is $+1.9 \times 10^{-6}$.

## 1. Introduction

Johnson noise is the electronic noise caused by thermally induced fluctuations in voltage and current that occurs in all electrical conductors at finite temperature. Although Einstein predicted the noise in his 1905 explanation of Brownian motion [1], it was more than two decades before it was measured by Johnson [2-3] and explained in detail by Nyquist [4]. Both Johnson noise and Brownian motion are examples of a fundamental thermodynamic relation called the fluctuation-dissipation theorem [5], which relates microscopic fluctuations to linear dissipative mechanisms.

For Johnson noise, the fluctuation-dissipation theorem describes the power spectral density of the noise voltage across the resistor [4-5]

$$S_R = 4hfR\left[\frac{1}{2} + \frac{1}{\exp(hf/kT)-1}\right], \qquad (1)$$

where $h$ is Planck's constant, $f$ is frequency, $R$ is the resistance, $k$ is Boltzmann's constant, and $T$ is the temperature of the resistor. Usually, Johnson noise is characterized by its mean-square voltage, conventionally called the noise power. For temperatures near 300 K and frequencies below 1 GHz, the noise power is approximated to better than 1 part in $10^9$ by Nyquist's law,

$$\overline{V_T^2} = 4kTR\Delta f, \qquad (2)$$

where $\Delta f$ is the bandwidth over which the noise is measured. Johnson noise is described as a "white noise" since the power spectral density $S_R = 4kTR$ is independent of frequency.

Because Johnson noise thermometers (JNT) are purely electronic, they offer an appealing alternative to the various forms of primary gas thermometry; constant volume gas thermometry [6], acoustic gas thermometry [7-11], and dielectric constant gas thermometry [12-13], all of which are limited by the non-ideal properties of real gases. Almost certainly, the numerical value assigned to Boltzmann's constant with the introduction of the 'New SI' [14] will be primarily determined by the values obtained by acoustic gas thermometry [15]. However, there remains the possibility of unknown systematic effects that might bias the gas thermometry results, and therefore an alternative determination using a different physical technique and different principles provides valuable assurance that any unrecognised systematic effects must be small. The CIPM's Consultative Committee on Thermometry (CCT) has proposed that the kelvin redefinition should proceed when the next CODATA adjustment assigns a value of $k$ with a relative uncertainty below $1\times10^{-6}$, supported by at least one determination from a second technique reporting a relative uncertainty below $3\times10^{-6}$ [16].

---

[1] This work is a contribution of the U.S. government and is not subject to U.S. copyright.

Historically, noise thermometry has not been a practical option for Boltzmann constant determinations because of three main challenges. Firstly, the noise voltages are extremely small, typically less than 2 μV rms. To practically measure such a signal, it has to be amplified by factors of $10^5$ or more while maintaining the relative accuracy of the signals at levels approaching $1\times10^{-6}$. Secondly, the random nature of the noise means that statistical uncertainty decreases as the square root of the number of samples, and data in excess of 20 TB are required to achieve relative uncertainties approaching $1\times10^{-6}$. This means integration times of many weeks and bandwidths well in excess of 100 kHz are required, and such wide bandwidths are not consistent with maintaining the required signal accuracy. Thirdly, with purely analog electronics, it is impossible to define the system bandwidth with the accuracy required.

Two technological breakthroughs changed the landscape. The first breakthrough came with the advent of sufficiently fast and accurate analog-to-digital converters (ADC) [17-18]. In combination with digital signal processing, fast ADCs have made it possible to process the signals in the frequency domain where bandwidths can be defined accurately. The second and most significant breakthrough was the development of ac-Josephson voltage synthesizers [19-20]. This made it possible to generate quantum-accurate pseudo-random noise with a power spectral density closely matched to that of the thermal noise, and hence operate the noise thermometer as a comparator rather than an amplifier and a meter, which dramatically reduced problems with linearity and accuracy of the electronics [21-25]. The quantum-accurate pseudo-random voltage-noise source (QVNS) also makes it possible to measure Boltzmann's constant directly in terms of fundamental constants, namely the Josephson and Planck constants [26].

NIST reported the first practical electronic measurement of the Boltzmann constant. This Johnson noise thermometer compared the noise powers from a thermal noise sensor at the water triple point (WTP) with the noise power synthesized by a QVNS [27]. The result, $k = 1.380651(17)\times10^{-23}$ J/K, is consistent with the 2010 CODATA value [28]. The $12.1\times10^{-6}$ combined relative uncertainty of this measurement was dominated by two terms; the "systematic effects that produce aberrations in the ratio spectra" ($10.4\times10^{-6}$) and statistical uncertainty due to the randomness of the noise ($5.2\times10^{-6}$). Neither the spectral aberrations nor the statistical uncertainty represent a fundamental limit of the measurement, but further improvements are necessary if noise thermometry is to make a useful contribution to the determination of $k$.

From 2010, we started a NIST/NIM collaboration project to develop another QVNS-JNT system at NIM, China. The NIM noise thermometer adopts the design pioneered by NIST [21-27, 29-33], with some variations [34-35]. This paper reports an improved Boltzmann constant determination by the NIM system with continuous improvements developed by NIM and NIST. In the following sections we describe the experimental apparatus, including the operating principles, and the design of the QVNS, the thermometer probe, and the correlator. This is followed by a description of the experimental algorithm and results, and a detailed review of the uncertainty analysis. Finally, we present the results and conclusions.

## 2. Experimental apparatus

### 2.1 Overview and operating principle

The most successful JNT technique for the medium- and high-temperature ranges is the switched-input correlator pioneered by Brixy for application in nuclear reactors [17, 36] and is now used routinely for most metrological noise thermometry. The switched correlator combines the amplifier-noise immunity of cross-correlators, first exploited by Fink [37], and the gain-instability immunity of the Dicke radiometer [38]. Digital noise thermometers also eliminate the analog multiplier traditionally employed in correlators by digitizing the signals from the two correlator channels and performing the multiplication and averaging functions in software.

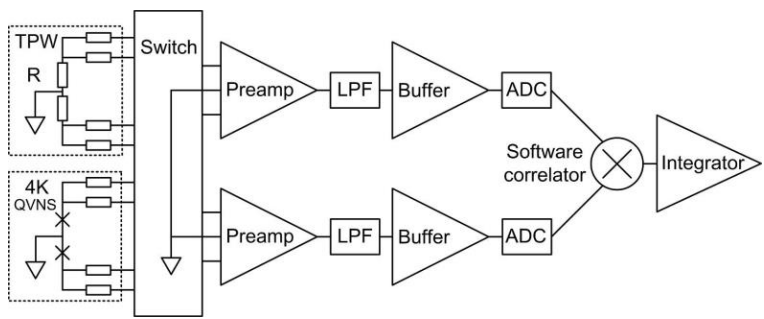

**Figure 1:** A simplified schematic diagram of the Johnson noise thermometer showing the thermal and QVNS noise sources, the switch, preamplifiers, low-pass filters (LPF) and ADCs. The cross-correlation and integration is performed in software.

Figure 1 shows a simplified schematic diagram of the QVNS-JNT. At the input there are two noise sources. The first is a resistor maintained in a TPW cell, producing Johnson noise with a power spectral density

$$S_R = 4kT_W X_R R_K, \qquad (3)$$

where $T_W$ is the temperature of the TPW, which currently defines the kelvin, and the sensing resistance is expressed as the ratio $X_R$ in units of the von Klitzing resistance $R_K \equiv h/e^2$ [39], where $e$ is the charge of the electron, and $h$ is Planck's constant. The second noise source is the quantum-accurate voltage-noise source (QVNS) [26-27], producing a pseudo-random noise for which the calculable power spectral density is

$$S_{Q\text{-calc}} = D^2 N_J^2 f_s M / K_J^2, \qquad (4)$$

where $K_J \equiv 2e/h$, is the Josephson constant, $f_s$ is a clock frequency, $M$ is the bit length of the digital code for the noise waveform, $D$ is an adjustable parameter of the software that generates the digital code that sets the amplitude of the synthesized QVNS waveform, and $N_J$ is the number of junctions in the Josephson array used in the QVNS. The JNT is normally operated with the power spectral densities closely matched, $S_{Q\text{-calc}} \approx S_R$.

The two correlator channels alternately amplify, filter, and sample the respective noise signals from the thermal and QVNS sources, which are then digitized by the ADCs and cross-correlated in software. The output of the correlator is proportional to the noise powers of the respective thermal and QVNS signals. Since the bandwidth of the system is defined digitally and is the same in the two configurations, the ratio of the noise powers gives the ratio of the power spectral densities $S_R / S_Q$. The Boltzmann constant is then determined by

$$k = h \frac{D^2 N_J^2 f_s M}{16 T_W X_R} \frac{S_R}{S_Q}. \qquad (5)$$

Contributions to the uncertainty in the determination of $k$ are therefore due to the uncertainties in Planck's constant, the QVNS clock frequency, the noise-power-ratio measurements, the realization of the triple point of water, and the resistance measurement. The remaining parameters in (5) are known exactly and contribute no uncertainty. By far the most significant sources of uncertainty are associated with the measurement of the ratio of the power spectral densities (details are given in Section 4).

**2.2 The thermal noise source**

The sensing resistor consists of two Ni–Cr-alloy foils on an alumina substrate, each with a nominal resistance value of 100 Ω. Two pairs of gold-coated leads on the hermetically sealed package are connected to the two ends of the series-connected resistors and a fifth lead is connected between the two resistors to form the ground connection (Figure 1).The resistor is mounted on a copper header soldered at the end of a 6 mm diameter and 0.1 mm thick thin-wall stainless-steel probe The probe is about 50 cm long, filled with argon gas and hermetically sealed. Two pairs of very thin coaxial cables, also of 50 cm length, transmit the differential noise signals from the resistor to the connectors at the top of the probe. The fifth lead is connected directly to the probe head and then to the aluminium box that shields the switch assembly (Figure 1). The probe is immersed in a TPW cell maintained in an ice bath contained within a stainless-steel Dewar. This probe design differs from the NIST design, which uses the same resistor package, but uses a pair of shielded twisted-pair copper leads and has the fifth wire connected to the shields for each twisted pair. The NIST resistor sense probe also uses flowing dry nitrogen gas and is inserted in a TPW cell whose temperature is maintained with a thermo-electric cooler [27].

**2.3 The QVNS**

The QVNS is a quantized delta-sigma (Δ-Σ) digital-to-analog converter (DAC) using oversampling techniques to produce a programmed sequence of high-speed pulses, typically with frequency of a few gigahertz. With appropriate algorithms and biasing [40], it produces a pseudo-random noise waveform with the desired harmonic content over a band from dc to several megahertz, well beyond the operating bandwidth of the JNT. The primary advantage of the QVNS is that each pulse from the DAC has a quantized area

$$\int V(t)dt = nh/2e \,, \tag{6}$$

where $n$ is an integer ($n = 1$ under normal operating conditions) [41]. This enables the synthesis of waveforms calculable exactly from the known sequence of pulses, the clock frequency of the pulse generator, and fundamental physical constants, according to (4).

The synthesis technique underlying the QVNS was originally developed for ac-Josephson voltage standards [19-20]. Conventionally, a high-speed pulse train, obtained through a software Δ-Σ modulator, is applied to the Josephson junctions to synthesize the desired voltage waveform, which contains the low-frequency component dominated by the expected waveform. The low-frequency component in the driven signal increases the system complexity and induces unwanted voltages through the on-chip inductances. Though the inductive error is negligible at frequency of a few kilohertz, it becomes significant at the higher frequencies desired for JNT measurements having larger bandwidth. Instead of the conventional synthesis method, we use a novel zero-composition method to synthesize the pseudo noise waveform [42]. The zero-compensation pulse pattern is obtained through two-level Δ-Σ modulation and then reconstructed by padding each positive pulse with two negative pulses of half amplitude for cancellation. This approach significantly reduces the low-frequency components of the resulting bipolar, ternary-level pulse train and thus the inductive voltage error [43].

In the QVNS chip that was fabricated at NIST in Boulder there are ten superconductor-normal metal-superconductor (SNS) Josephson junctions ($N_J = 10$) in each of two arrays [40]. The critical current of the Josephson junctions is around 6 mA and the characteristic frequency is about 5 GHz. The chip is mounted on a flexible package in a magnetically shielded probe. The probe is cooled to 4 K in a 100 l liquid-helium Dewar. The spectrum of the QVNS waveform is composed of a series of odd harmonic tones at multiples of the pattern repetition frequency, $f_1 = f_s/M = 90$ Hz, up to 4 MHz. All tones have the same amplitude but random phase (see Figure 2). When the QVNS is used to measure $k$, the rms voltage amplitude of each of the tones is set to be 23.3034565 nV, so that the synthesized waveform's average power spectral density matches the thermal noise power spectral density to about 0.01 %.

In addition to providing the link to Planck's constant, the QVNS has important advantages that have enabled significant improvements in the bandwidth and accuracy of the noise thermometer. Unlike a resistor noise source, the QVNS output voltage is inherently independent of its output impedance. This makes it possible to match both the noise powers and the output impedances of the thermal and QVNS sources, overcoming the matching conflict inherent in conventional Johnson noise thermometers that use two thermal noise sources [29]. To achieve the impedance match, four resistors terminate the QVNS transmission line such that one resistor is placed on-chip in each of the four QVNS output leads. Since these resistors are at 4 K, they produce only a small amount of uncorrelated noise. There are also four small-valued resistors placed in each lead of the sensing-resistor transmission line. These resistors ensure the thermal and QVNS sources have both the same uncorrelated noise powers to minimize effects of any non-linearity. To ensure a good match of the frequency responses of the thermal and QVNS probes across a wide bandwidth, small parallel-connected capacitors and series-connected inductors with carefully trimmed values that are inserted at the input of the switch circuit. Close matching of the frequency responses of the two probes and their transmission lines minimises the 'spectral mismatch' error and ensures that the JNT can be operated over a greater bandwidth and with a reduced measurement time.

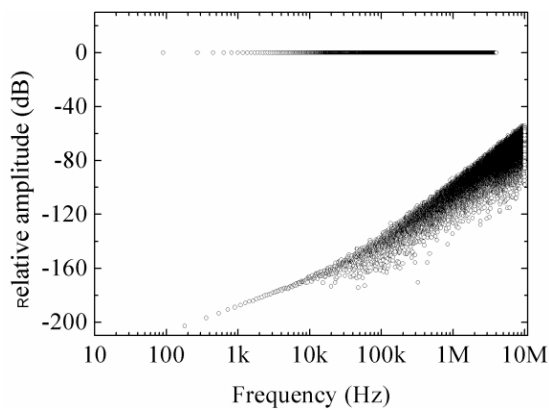

**Figure 2:** The first 10 MHz of the QVNS spectrum calculated from the QVNS code sequence. The upper branch shows the odd harmonics up to 4 MHz. The lower branch shows the even harmonics up to 4 MHz, and all harmonics above 4 MHz. Below 4 MHz, the amplitudes of the even harmonics are indicative of the accuracy of the amplitudes of the odd harmonics.

**2.4 The correlator**

The respective noise powers of the resistor and the QVNS are alternately measured by the correlator, depending on the relay configuration of the switching network, as indicated in Figure 1. Each channel of the correlator is composed of a low-noise preamplifier and gain blocks with a total gain of about 70 dB, filters that define the overall measurement bandwidth and prevent aliasing, and a fast ADC.

The switching circuit consists of four latching relays mounted on a FR4 printed circuit board (PCB). The relays are controlled by a field-programmable gate array (FPGA) to alternate between the noise sources. The power supply of the FPGA is turned off during the measurement and controlled via optical fiber so as to eliminate the FPGA as a source of electromagnetic interference close to the input signal leads. On the PCB, the signal traces are symmetrically placed in the top layer and the total length from input lead to output lead is about 15 mm. The copper ground plane is in the bottom layer, such that the area opposite to the signal traces is removed. This design minimizes the capacitance to ground of the signal leads and hence also minimizes the effect of dielectric losses [27].

The preamplifier, based on a common-source-common-base FET-bipolar cascode input stage operated without feedback [44], has been designed for lower noise, higher bandwidth and higher common-mode rejection [30]. These features are necessary to meet the demanding JNT requirements of low input noise voltage, very low input noise current, low noise-current-noise-voltage correlation, high input resistance, low input capacitance, and high common-mode rejection ratio. The filters are passive *LC*-ladder filters implementing an 11$^{th}$-order Butterworth response with a cut-off frequency of 1.8 MHz. Two filters are placed in each amplifier chain to ensure that the contribution of aliased signals to the measured noise power is negligible.

Custom-made 16-bit-ADC boards sample the signals in each channel for 1 s periods with a sampling frequency of 4 MHz. The ADCs are clocked and triggered externally via optical fibres, with the phases of the clocks carefully adjusted so that the acquisitions of the two correlator channels are synchronized. Although the ADC is only 16 bits, dither due to both correlated and uncorrelated noise ensure that quantisation effects and differential non-linearity have no practical effect on the measurement [45]. Large-scale and integral non-linearity effects are still a factor and these are discussed in Section 4.1.3.

Once each signal has been sampled for 1 s, fast Fourier transforms (FFT) of the signals are computed yielding complex spectra with 1 Hz frequency-resolved FFT bins and a 2 MHz Nyquist frequency. To ensure that the QVNS tones are located in a single FFT bin, the ADC clocks are locked to the same frequency reference as the QVNS clock. In addition to the FFTs, the computer carries out a complex frequency-domain cross correlation of the FFT spectra for the two channels, which reduces the uncorrelated amplifier noise voltages. The autocorrelation for each channel is also calculated for diagnostic purposes. All of these spectra are then accumulated for 100 s into an average spectrum to provide a compact form of data storage, defined as one 'chop', for post processing and for final computation and diagnostic checks. These computations are carried out in real time, and every 100 s, the correlator is switched ('chopped') between the resistor and the QVNS.

## 2.5 Shielding and grounding

Very good shielding and grounding are necessary because electromagnetic interference (EMI) adversely affects the measurements by inducing preamplifier overloads and by creating systematic errors due to coupling into both correlator channels [33]. The switching circuit, the amplifier chains and the lithium-ion batteries powering the analog parts of the correlator are mounted in separate aluminium boxes. The whole of the measurement electronics is then placed in a mu-metal box. The signal connections between the electronics box and the QVNS and the resistor probe are heavily shielded and use separately shielded twisted pairs to minimize any EMI and crosstalk between the channels. To eliminate ground loops, only the ground of the pulse generator that drives the Josephson junctions through microwave cables is directly connected to the earth point via a low-resistance connection.

The whole of the JNT system is housed in a large underground screened room that has a filtered mains supply and fully isolated communications systems. The room is also deliberately located away from major magnetic machinery such as transformers and electric motors, to reduce the effects of low-frequency magnetic EMI. With all of these precautions, no EMI signals are observed in the integrated spectra for both thermal and QVNS noise measurements.

## 3. Measurement result

Currently, two different operating methods for long-term measurements are being used. At NIST, the measurement time is limited only by the He-boil-off in the storage Dewar used to cool the QVNS circuit. The batteries of the JNT electronics are automatically switched and recharged and the triple-point cell is continuously cooled by a Peltier cooler, allowing for measurements of more than 200 hours. Figure 3 shows an example for a 146 hour continuous measurement of 2439 chops for a filter bandwidth of 800 kHz. The shielded system is located in a typical non-shielded laboratory. The statistical uncertainty still shows no saturation tendency and the fitting residual shows no frequency dependence up to about 800 kHz for 487,800 one-second averages.

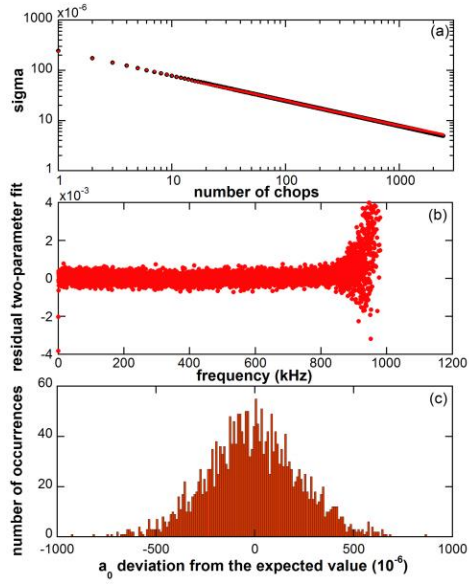

**Figure 3:** An example of a 135.5 hour continuous measurement of 2439 chops showing (a) the statistical uncertainty, (b) the frequency dependence of the residual for a two-parameter fit, and (c) the error distribution of the $a_0$-diviation.

At NIM, individual measurements are performed for about 20 hours each, limited by the capacity of the batteries for the digitizers and the maintenance period of the triple point of water cell in the ice bath. For the results reported here, 45 such measurements were carried out and the results are combined to reduce the statistical uncertainty. The resistance value of the sense resistor was checked before and after every measurement with a DC resistance bridge.

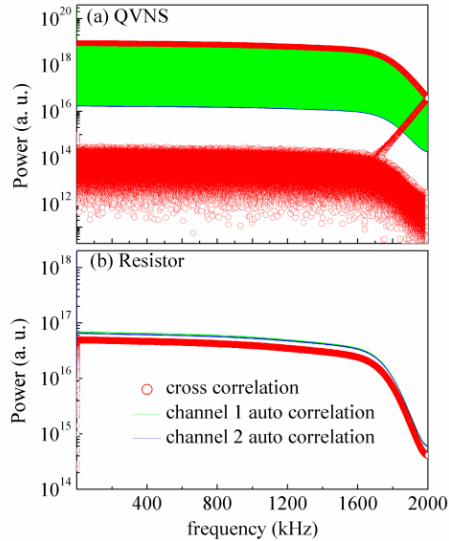

**Figure 4:** Measured FFT spectra from noise sources of (a) the QVNS and (b) the sense resistor. The blue and green lines are the auto-correlation spectra for each of the two amplifier channels, and the red circles represent the cross-correlation spectra between channels.

The accumulated auto-correlation and cross-correlation spectra for one measurement with 380 chops are shown in Figure 4. Both the QVNS and thermal noise spectra are very clean such that they show no evidence of EMI signals in the 2 MHz measurement bandwidth. Note also that the amplitude of the spectra decreases visibly above 1.6 MHz due to the frequency response of the LPF. The QVNS spectrum shows that the aliased part of the spectrum is quickly attenuated such that it is already 60 dB lower than the in-band signal at 1.6 MHz.

Once the measurements are complete, all of the cross-correlation and auto-correlation spectra, 13982 chops data for each of the thermal and QVNS sources, are averaged respectively, and the real part of each of the thermal and QVNS cross-spectra are reduced in resolution by summing the 1800 FFT bins in each 'block'. This rebinning is necessary because most of the bins in the QVNS spectrum are largely empty due to the absence of the tones, and a direct ratio of the QVNS and thermal spectra would not have similar ratios in each bin. The ratio of the rebinned thermal and QVNS spectra is then computed. Forming the ratio spectrum is not only essential for computing $S_R/S_Q$ in (5), it also removes the amplifier and analog filter frequency responses from the calculations, and ensures equal weighting of all spectral elements in the calculation. The equal weighting has the effect of maximizing the correlation bandwidth of the thermometer [32] and yields a small reduction in the statistical uncertainty. Note, particularly, that all of the spectral averages are computed before the noise-power ratio spectrum is calculated. This calculation sequence is necessary to avoid a bias caused by the (non-linear) division calculations operating on the stochastic data [46-47].

The noise-power ratio spectra calculated with the rebinned thermal noise and QVNS spectra are shown in Figure 5. Figure 5a shows the relative differences in the auto-correlation spectra, which include noise from the preamplifier and other uncorrelated noises, and the cross-correlation spectrum from which the noise-power-ratio spectrum is computed. All three plots show the differences relative to the corresponding thermal spectrum. It can be seen that the correlated noise powers are well matched, and within 1 part in $10^3$ up to about 800 kHz. The uncorrelated noise powers are not quite so well matched, but the effect of mismatch here is not so important (see section 4). Figure 5b shows the ratio of the real parts of the accumulated thermal and QVNS spectra, which again shows the close match of the frequency responses for the two signals.

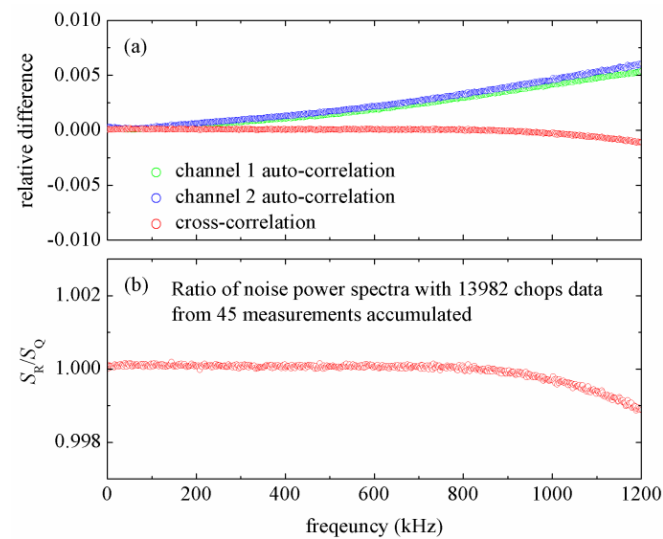

**Figure 5:** The averaged spectra resulting from the 45 measurements. (a): the relative differences in the auto-correlation and cross correlation spectra, (b) the ratio of the thermal and QVNS cross-correlation spectra.

In spite of the closely matched noise powers and transmission line impedances, small differences remain and result in slightly different transfer functions for the thermal and QVNS signals. Furthermore, there remain errors due to the amplifier noise currents. When connected to the thermal source, the preamplifier noise currents pass through the sense resistor and are measured by both correlator channels, leading to extra undesirable correlated noise power and a systematic error. This error is absent when the preamplifiers are connected to the QVNS, which acts as a short circuit for the noise currents.

These errors, although relatively small, result in deviation of the noise-power ratio from unity, especially at higher frequency. Both errors are expected to have $f^2$ dependence, and in the previous determination of $k$ [27], the resulting quadratic frequency response of the noise-power ratio was corrected by least-squares fitting of the ratio spectrum with a two-parameter model, $a_0+a_2f^2$, where the coefficients $a_0$ of various measurement are used to calculate $k$, and the associated standard deviations from the fits determined the relative statistical uncertainty. Nevertheless, there remained small unexplained spectral aberrations manifest as variations in the values of the fitted parameter values when the fits were performed over different bandwidths. These effects of spectral aberrations dominated the final measurement uncertainty [27]. For the data and analysis presented here, a higher-order model is employed to correct the residual spectral aberration, and described in detail in Section 4.1.

Recent work at NIST suggests that some spectral aberrations can be attributed to ferrite-cored inductors in the input circuits in front of the preamplifier, which may be used to suppress the common-mode Colpitt's oscillation that can occur from interaction between the differential FET input stages during the sense resistor measurement. A comparison of the typical bandwidth dependence of $a_0$ for different ferrites is shown in Figure 6. The ferrites can introduce losses and non-linearities that may arise at lower frequencies. Although the inductors are connected in common mode, there is sufficient differential-mode leakage inductance for the cores to introduce frequency dependence. A more detailed analysis will be published elsewhere.

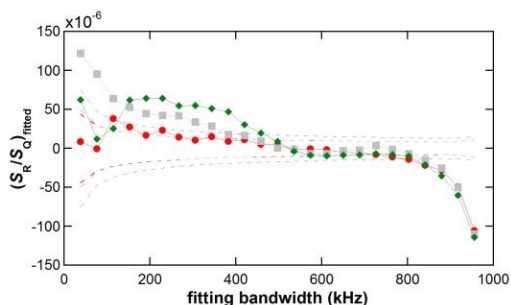

**Figure 6:** Fitted ratio $S_R/S_Q$ as a function of fitting bandwidth for ferrite beads with different specifications and placed in the input circuit before the preamplifier.

## 4. Sources of error and uncertainty

Because all of the variables in Equation (5) for the Boltzmann constant appear as multiplying or dividing factors, the propagation-of-uncertainty expression for the uncertainty in $k$ is most simply expressed in terms of relative uncertainties:

$$\frac{u^2(k)}{k^2} = \frac{u^2(h)}{h^2} + \frac{u^2(f_s)}{f_s^2} + \frac{u^2(T_{\text{TPW}})}{T_{\text{TPW}}^2} + \frac{u^2(X_R)}{X_R^2} + \frac{u^2(S_R)}{S_R^2} + \frac{u^2(S_Q)}{S_Q^2}, \qquad (7)$$

where each relative uncertainty can be expressed in parts per million (ppm). Note that there are no uncertainty terms in $D$, $N_J$ or $M$ because they are exact integers or numbers.

The contributions occur in four main groups: 1) those associated with the QVNS waveform generation, 2) those associated with the realization of the water-triple-point temperature, 3) those influencing the measurement of the sensor resistance in terms of $R_K$, and 4) those associated with the measurement of the ratio of the power spectral densities. All of these uncertainty contributions are discussed in detail in the following sections.

4.1 The power spectral ratio

*4.1.1 Statistical uncertainty*

Because the thermal noise is random, it results in a purely statistical uncertainty that makes the largest contribution to the total uncertainty. The two measurement phases; one for the measurement of the thermal noise power and one for the QVNS power measurement, have slightly different contributions [32]:

$$\frac{u^2(S_R)}{S_R^2} = \frac{1}{2\tau_R \Delta f_c}\left[\left(1+\frac{S_{n1}}{S_R}\right)\left(1+\frac{S_{n2}}{S_R}\right)+1\right] \qquad (8)$$

and

$$\frac{u^2(S_Q)}{S_Q^2} = \frac{1}{2\tau_Q \Delta f_c}\left[\left(1+\frac{S_{n1}}{S_Q}\right)\left(1+\frac{S_{n2}}{S_Q}\right)-1\right], \qquad (9)$$

where $\tau_Q$ and $\tau_R$ are the integration times for the thermal and QVNS measurement phases respectively, $\Delta f_c$ is the correlation bandwidth, and $S_{n1}$ and $S_{n2}$ are the power spectral densities of the uncorrelated equivalent input noise voltages in the two channels of the correlator. It is assumed that the various power spectral densities are independent of frequency. Because the noise-power ratios are measured in the frequency domain, the correlation bandwidth is simply defined by the FFT bins selected for inclusion in the measurements. Note that $\tau_R \Delta f_c = \tau_Q \Delta f_c$ is the number of selected FFT bins, and the uncertainty decreases as $1/\sqrt{\tau_R \Delta f_c}$, as expected for random noise.

The uncertainty contributions of the two measurement phases are different, as evident from the differing signs of the last term in each of (8) and (9). The sign is positive for the thermal measurement and negative for the QVNS measurement. The negative sign for the QVNS measurement arises because the QVNS signal is not truly random but has a constant noise power when integrated over multiples of the code recycle time, and hence, the uncertainty is lower than for the thermal noise-power measurement. The uncertainty in the QVNS measurement is zero when the uncorrelated noise, largely due to the preamplifier equivalent in put noise voltage, is zero.

Because the two measurement phases make different contributions to the uncertainty, it is possible to minimise the total uncertainty by spending a greater fraction of the measurement time integrating the thermal noise signal [32]. However, the advantage is modest and for the measurements reported here, the integration times were the same for both phases. Since the total measurement time, $\tau$, is the sum of the integration times for the thermal and QVNS measurements, and $S_Q \approx S_R$, the statistical uncertainty is approximately

$$\frac{u^2(S_R)}{S_R^2} + \frac{u^2(S_Q)}{S_Q^2} \approx \frac{2}{\tau \Delta f_c}\left(1+\frac{S_{n1}}{S_R}\right)\left(1+\frac{S_{n2}}{S_R}\right). \qquad (10)$$

However, this result applies only when every selected QVNS-tone block in the ratio spectrum contributes equally to the estimate of $S_R/S_Q$. The use of the least-squares fit to correct for any frequency dependence in the power spectra (see discussion below) changes the weighting of the different blocks, and the statistical uncertainty increases.

*4.1.2 Spectral aberrations*
Because the same correlator is used to measure the noise power for the thermal and QVNS signals, the ratio spectrum is independent of the gain and frequency response of the correlator. Frequent (~ 100 s) switching between the two sources also eliminates the effects of drifts in gain and frequency response. However, because the response of the transmission lines connecting the sensor and QVNS to the correlator are independent, small differences in the frequency responses of these components do have an impact on the measured ratio spectrum.

There are two main causes for spectral mismatches. Firstly, small losses (i.e. finite loss tangents

tan$d$) caused by imperfect dielectrics associated with various shunt capacitances (in the components used to assemble either the QVNS, the thermal sensor, or both), lead to contributions in the transfer functions for R and QVNS that are linear in $f$ tan$d$, and differences in such terms between the R and QVNS lines can then be manifest in the ratio spectrum [48]. In the previous measurement [27], the relative uncertainty due to this effect was estimated to be about $2\times10^{-6}$. With this in mind, every effort has been made to minimise the lengths of PCB tracks in the input circuits and to minimise the use of fiberglass PCB, which is a major contributor of this effect. The net effect can be simulated, given reasonable estimates for the various impedances, but this modelling has not been completed. It is expected that any error in the present measurement caused by dielectric loss is less than $1\times10^{-6}$.

The second and largest source of spectral aberrations arises from mismatches between the source impedances (thermal and QVNS) and the load impedance presented to each source by their respective transmission lines connecting them to the preamplifiers, and the input impedances of the preamplifiers. Ideally, the very small attenuations that occur should be the same for both sensors so that the ratio of two spectra is flat. When the transmission lines have been carefully matched, any remaining spectral mismatches occur only at high frequencies, and ideally the ratio spectra have the appearance of the response of a high-order Butterworth filter, which is maximally flat throughout its passband. Figure 4, which shows the measured ratio spectrum, exhibits these features. For frequencies up to about 800 kHz, the ratio spectrum is flat, and then above 800 kHz the ratio begins to fall rapidly.

The ratio spectrum, including residual mismatches, can be modelled by

$$R(f) = \frac{S_R}{S_Q}\left(1 + a_2 f^2 + a_4 f^4 + a_6 f^6 + ...\right), \qquad (11)$$

where $S_R/S_Q$ is the desired low-frequency limiting value of the power-spectral ratio required for the measurement of $k$ (equation 5), and the coefficients $a_2$, $a_4$, $a_6$ represent the effects of the various impedance mismatches, and are ideally very small. Only even order terms are included in the series expansion because it is assumed that the transmission lines are short enough to be modelled by lumped inductances and capacitances, which would give rise only to even order terms in the real parts of the power spectra.

There are two approaches for managing the spectral mismatches. The first is to limit the bandwidth so that the mismatch effects are small, and this was the approach adopted with most conventional noise thermometers prior to the NIST measurement. With this approach, $S_R/S_Q$ is measured simply as the average of the ratio spectrum between zero frequency and the chosen correlation bandwidth, $\Delta f_c$,

$$\left.\frac{S_R}{S_Q}\right|_{meas} = \frac{S_R}{S_Q}\left(1 + \frac{a_2}{3}\Delta f_c^2 + \frac{a_4}{5}\Delta f_c^4 + \frac{a_6}{7}\Delta f_c^6 + ...\right), \qquad (12)$$

so that the measured value of $S_R/S_Q$ is biased by the frequency dependent terms. Typically, when the measured value of $S_R/S_Q$ is plotted against frequency, the bias will increase rapidly and emerge from the noise, very much as shown in Figure 4.

As is well known, if the spectral ratio alone (a constant) is fitted to the data, the uncertainty in the fitted value of $S_R/S_Q$ should fall in proportion to the square root of the number of data points

$$u(S_R/S_Q) = \frac{\sigma}{\sqrt{N}}, \qquad (13)$$

where $\sigma$ is the variance in the spectral ratio for each QVNS-tone block, and $N$ is the number of QVNS-tone blocks included in the fit. Operation of the noise thermometer therefore requires a compromise between the statistical uncertainty, which decreases as the square root of the bandwidth, and the uncertainty due to spectral mismatches, which increases according to a power law, according to (12).

The second approach is to apply a least-squares fit of (11) to the measured ratio spectrum, so that values are determined for all of the parameters in the model, including the desired low-frequency power spectral ratio. The cost of this approach is that the uncertainty in the fitted value of $S_R/S_Q$ increases with the number of fitted parameters. Suppose that the least-squares fit estimates the low-frequency power spectral ratio $S_R/S_Q$ by fitting the function

$$R(f_i) = \frac{S_R}{S_Q}\left(1 + a f_i^2\right) \qquad (14)$$

to the ratio spectrum, where $f_i$ are the frequencies at the centre of each QVNS tone block, and hence the low-frequency power spectral ratio is estimated as

$$\frac{S_R}{S_Q} = \frac{\sum f_i^4 \sum R(f_i) - \sum f_i^2 \sum f_i^2 R(f_i)}{N \sum f_i^4 - \left(\sum f_i^2\right)^2}. \qquad (15)$$

Since, $i$ ranges from 1 to $N$ and $f_i = f_1(2i-1)$ where $f_1$ is the QVNS-tone pattern repetition frequency, equation (15) identifies the sensitivity coefficients for $S_R/S_Q$ with respect to each $R(f_i)$ as

$$\frac{d(S_R/S_Q)}{dR(f_i)} = \frac{\sum f_i^4 - f_i^2 \sum f_i^2}{N \sum f_i^4 - \left(\sum f_i^2\right)^2}. \qquad (16)$$

The total uncertainty in $S_R/S_Q$ is proportional to the quadrature sum of the sensitivity coefficients (assuming that the uncertainty for the ratio in each block is independent of frequency) the uncertainty in the fitted value of $S_R/S_Q$ is

$$\sum_{i=1}^{N}\left(\frac{d(S_R/S_Q)}{dR(f_i)}\right)^2 \sigma^2 = \frac{\sigma^2}{N} \frac{3}{16} \frac{(12N^2 - 7)}{(N^2 - 1)} \approx \frac{\sigma^2}{N} \frac{9}{4}. \qquad (17)$$

so for large $N$, the variance in $S_R/S_Q$ determined by a second-order least-squares fit is larger than $1/N$ by the factor 9/4. If the fit is extended to the fourth- and sixth-order terms, the variances increase by the factors 225/64 and 1225/256 respectively. As the order of the fit increases, the uncertainty in the calculated value of $S_R/S_Q$ increases. The least-squares fit therefore adds an additional complication in the compromise between bandwidth, uncertainty, and the effects of the spectral mismatches.

The NIST thermometer was the first to combine the two approaches by using a second-order least squares fit combined with a bandwidth restriction. With the second order least-squares fit applied to the ratio spectrum, the bias in the measured spectral ratio is

$$\left.\frac{S_R}{S_Q}\right|_{meas} = \frac{S_R}{S_Q}\left(1 - \frac{3a_4}{35}\Delta f_c^4 - \frac{2a_6}{21}\Delta f_c^6 - \frac{a_8}{11}\Delta f_c^8 ...\right). \qquad (18)$$

Least-squares fits using higher-order models produce similar equations where the terms in the expressions for the bias omit the terms included in the fit.

There remains the problem of determining the best compromise between bandwidth and complexity of the fit. Figure 7 below shows the fitted value for $S_R/S_Q$ (the thick lines with marked points) as a function of correlation bandwidth and the order of the least-squares fit for the data reported here. The fits included in the analysis are zeroth-, second-, fourth-, sixth- and eighth-order. The data are all plotted as deviations from the point representing the value of $S_R/S_Q$ for the fourth-order fit and a bandwidth of 550 kHz. The envelopes formed by matching pairs of dotted lines in Figure 5 show the relative statistical uncertainty measured for the corresponding fits.

Ideally, each of the $S_R/S_Q$ curves shown in the figure should exhibit a characteristic shape. At low frequencies, the points should follow a type of random walk that lies, more or less, within the corresponding uncertainty envelope, and converge slowly (as $\Delta f_c^{-1/2}$) towards the true, but unknown, value of $S_R/S_Q$. At some higher frequency, each of the curves should rapidly diverge as the $\Delta f_c^n$ effects of the spectral mismatches become apparent. All of the curves in Figure 7 exhibit this shape to some degree. Also shown in Figure 7 is a rectangular box enclosing a consistent set of data points from all but the zeroth-order fit. As expected, the low-order fits converge on the box at lowest frequencies before diverging, while the higher-order fits converge at higher frequencies. Amongst these points, the one that has the lowest statistical uncertainty and does not sit close to the part of the curve that diverges rapidly, is the point for fourth-order fit and 550 kHz bandwidth. The uncertainty in the fitted value for $S_R/S_Q$ for this point is $3.2\times10^{-6}$.

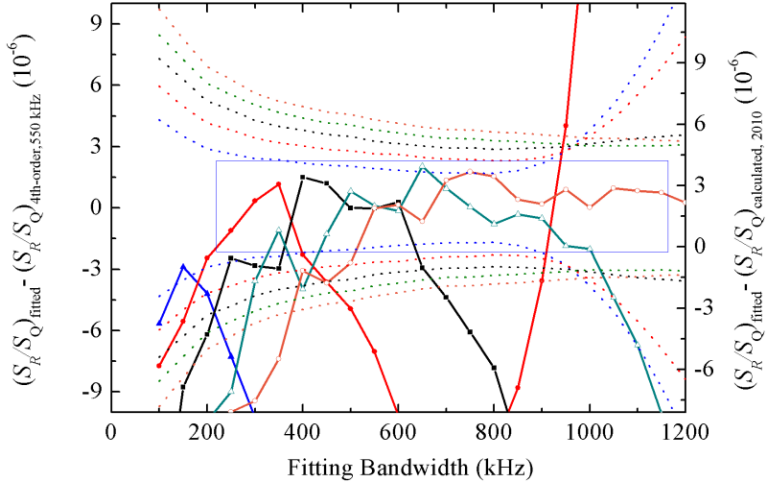

**Figure 7:** Plot of the estimate of $S_R/S_Q$ versus bandwidth and the order of the least-square fit: blue (zeroth-order), red (second-order), black (fourth-order), green (sixth-order), brown (eighth-order). The dotted lines show respective measured statistical standard uncertainty envelopes.

There are two aspects of the uncertainty envelopes in Figure 7 worthy of further discussion. Firstly, none of the measured envelopes follow the expected $\Delta f_c^{-1/2}$ dependence. At high frequencies especially, the residuals in the fit include the effects of the spectral mismatches as well as the effects of the random noise. The effect of the spectral mismatches is most apparent in the uncertainty envelopes for the zeroth- and second-order fits where they diverge rapidly above 800 kHz. Secondly, the uncertainty envelopes in Figure 5 describe the uncertainty in the fitted value of $S_R/S_Q$, but they do not describe the extent of the random walk of the fitted values because they do not account for the correlations between adjacent data points. For example, the data used for the 500 kHz fits includes 100% of the data used for all the fits for smaller bandwidths. When the effects of the correlations are included in the zeroth-order fit, the envelopes are modified by the factor $\left|1-\Delta f_c/\Delta f_{c,\text{ref}}\right|^{1/2}$ where $\Delta f_{c,\text{ref}}$ is the bandwidth for the point to which the plots are referenced.

*4.1.3 Non-linearity*
In order to amplify the ~ 2 µV rms signals to a level appropriate for the ADCs, the gain in each channel of the correlator must be about ~$10^5$. Inevitably, the non-linearity in the various amplifier stages, filters, and ADCs accumulates to introduce a significant error in the noise-power measurement.

The effects of non-linearity are managed by operating the correlator so that the effects on both the thermal and QVNS signals are, ideally, identical, and therefore the ratios of the noise powers used to compute $S_R/S_Q$ are unaffected by the non-linearity. There are three main operating requirements to ensure the effects of non-linearity are minimised [29]. Firstly, the average power spectral densities of the QVNS and thermal sources must be the same to ensure that the total noise powers are the same in the two measurement phases. The match is achieved by adjusting the amplitude of the QVNS signal. Secondly, the QVNS must produce a signal that closely approximates the Gaussian distribution of the thermal noise voltage. This ensures that all of the various distortion products (effectively the moments of the distribution) are the same for both the thermal and QVNS signals. For a pseudo-random noise signal composed of a frequency comb with $m$ sinusoids of constant amplitude and random phase, the moments of the frequency comb converge to the Gaussian moments as $1/m$. Thus, for a noise thermometer with, say, 100 ppm distortion, more than 22,200 sinusoids used in the QVNS waveform are sufficient to ensure that the differences in distortion products are below 0.025 ppm of the measured signal. Thirdly, because some of the distortion products involve the uncorrelated preamplifier noise voltages, the uncorrelated noise powers in each channel must also be the same for the two measurements.

While carefully matched operating conditions ensure that the effects of non-linearity are small, there remains the problem of knowing how well the matching conditions are met, and of estimating the uncertainty due to the remaining mismatches. A study of the various causes and effects of non-linearity [49] concluded that the causes were so numerous, and the effects so variable, that a physical model of the non-linearity was not practical. Instead, any model of the non-linearity must be empirical and based on measurements made with deliberate mismatches of the noise power. One of the advantages of a noise thermometer operating a QVNS is that it is possible to change the QVNS voltage without changing any of the other operating conditions of the JNT, and this makes it possible to measure the non-linearity.

Suppose the (voltage) transfer function for each channel of the correlator, when referred to the correlator input, is represented as the Taylor series [29]

$$V_s = \sum_{t=0}^{\infty} a_{st}(V + V_{n,s})^t, \quad a_{s1} = 1, \tag{19}$$

where $V$ is the voltage to be measured, the index $s$ indicates the channel number ($s = 1, 2$), the index $t$ indicates the polynomial order of the distortion coefficient, and $V_{n,s}$ is the equivalent input noise voltage for channel $s$. The coefficients $a_{s0}$ therefore represent the offset voltages, and $a_{s1}$ represent the linear gains and are assumed to be equal to 1 for the purpose of analysis. All of the $a_{st}$ except $a_{s1}$ represent unwanted nonlinear terms, and are assumed to be small.

When the two Taylor series are multiplied and averages calculated, the most significant terms in measurements of noise power arise from the third order non-linearities:

$$\overline{V_{\text{meas}}^2} \approx \overline{V^2} + 3(a_{13} + a_{23})\left(\overline{V^2}\right)^2 + 3\left[a_{13}\overline{V_{n1}^2} + a_{23}\overline{V_{n2}^2}\right]\overline{V^2}. \tag{20}$$

The ratio of the QVNS and thermal noise-power measurements is therefore

$$\frac{\overline{V_{R,\text{meas}}^2}}{\overline{V_{Q,\text{meas}}^2}} \approx \frac{\overline{V_R^2}}{\overline{V_Q^2}}\left[1 + 3(a_{13} + a_{23})\left(\overline{V_R^2} - \overline{V_Q^2}\right) + 3a_{13}\left(\overline{V_{n1,R}^2} - \overline{V_{n1,Q}^2}\right) + 3a_{23}\left(\overline{V_{n2,R}^2} - \overline{V_{n2,Q}^2}\right)\right], \tag{21}$$

which clearly shows the requirement for matching the various noise powers. The equation also shows that the non-linearity in the noise power ratio should be proportional to the noise-power mismatch.

To measure the effects of the non-linearity, the Boltzmann constant measurement was repeated with several deliberately mismatched values of $S_R/S_Q$. The additional measurements were run only for a single day, so the uncertainties are much larger than that for the primary determination. Figure 8 below shows the variations in the measured value of $k$ versus the degree of mismatch. The line

$$k_{\text{meas}} = k\left[1 + \varepsilon(S_R/S_Q - 1)\right], \qquad (22)$$

which assumes that the amplifier noises are matched, was fitted to the data and the non-linearity coefficient was determined to be $\varepsilon = (39 \pm 20) \times 10^{-6}$. Since the measured ratio of $S_R/S_Q$ used for the determination deviates from unity by less than 0.0002, the correction is negligible. A total uncertainty of $0.1 \times 10^{-6}$ is assigned for non-linearity effects due to both the correlated noise-power mismatch and the uncorrelated noise-power mismatch, with no correction performed.

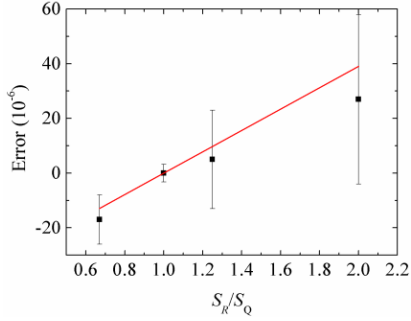

**Figure 8:** The non-linearity of the noise thermometer versus $S_R/S_Q$. The red line indicates the best fit to the measured points (black squares).

*3.1.4 Electromagnetic interference (EMI)*
EMI is an insidious source of error. It is typically intermittent, caused by nearby electrical machinery not associated with the JNT, and for magnetically-coupled EMI, very difficult to shield [49-50]. While statistical tests on the averaged power spectra can readily detect stationary single-frequency EMI, in general spectral tests are not sufficiently powerful to detect all types of EMI [51].

Evidence of the absence of EMI effects in the QVNS measurements was obtained by operating the QVNS so that it generates zero volts. Any non-zero noise power then indicates an error due to EMI. A similar test was carried out with the thermal noise source but this requires a 'dummy' thermal sensor that forms a four-terminal zero of resistance (so that it generates zero correlated noise), and has the same geometric layout as the real thermal sensor, but generates no correlated noise [White and Mason]. After a full day of integration, these tests yielded residual correlated-noise powers of relative magnitude less than $0.2 \times 10^{-6}$ in both the QVNS and resistor probes. We estimate a relative standard uncertainty of $0.4 \times 10^{-6}$ contributed by the possible EMI effect.

4.2 QVNS Waveform
The quantized nature of the voltages produced by the QVNS and its very wide bandwidth ensure that the uncertainties arising from the QVNS are small. The most significant contribution comes from the quantisation noise due to the generation of a continuous baseband signal from an integer number of high-frequency QVNS pulses. The software generating the code for the QVNS exploits first-order noise shaping to push the quantization error to the high-frequency end of the spectrum (see Figure 2). The resulting error integrated over the bandwidth used in the JNT contributes about $1 \times 10^{-7}$ to the relative uncertainty. Other, negligible sources of uncertainty include the uncertainty in the reference frequency and in Planck's constant [28].

4.3 Triple point of water

*4.3.1 Triple point realisation*
The TPW cell used in present measurements is made of borosilicate glass and contains distilled and de-gassed water. The diameter of the thermometer well is 14 mm and the height of the ice mantle above the thermal noise sensor is about 23 cm. By definition, the water-triple point temperature is 273.16 K exactly. However, there are errors and uncertainties associated with the practical realization of the triple point, all of which are summarised in detail in [52]. Because the cell was manufactured without an isotopic analysis, and chemical analysis would require the destruction of the cell, the temperature realized in the cell was determined by comparing it to the NIM standard TPW cell with a long-stem standard platinum resistance thermometer (SPRT). The temperature realized in the cell was found to be 1.46 K higher than that in the NIM standard cell. Such a temperature difference is peculiar, but could arise from a high heavy-isotope fraction or chemical impurities introduced with the cleaning of the glass. Though the reason is not clear yet, an extended comparison with the NIM standard cell demonstrated that the cell temperature is very stable and repeatable.

Water-triple-point measurements also require a correction due to the hydrostatic pressure of the water at the level of the SPRT or thermal noise sensor, which is typically 0.007 mK/cm. The corrected temperature used in the calculations is 273.1613 K. The associated uncertainties include 0.08 mK for the temperature realized by the cell and uncertainty of comparison NIM standard TPW cell [53], 0.02 mK uncertainty from the hydrostatic pressure correction, the 0.01 mK standard deviation of the comparison measurement with the SPRT.

*4.3.2 Immersion effects*
Triple point of water cells are designed for use with standard platinum resistance thermometers, the glass sheaths of which are poor thermal conductors and therefore do not conduct heat into the cell. The immersion characteristics for the thermal JNT probe are not so good, so potentially give rise to immersion errors.

To assess possible immersion effects a replica probe with a small platinum thermometer sensor was used to investigate the immersion characteristics of the probe. The measurement result is shown in Figure 7. The immersion error is expected to decay exponentially with the increasing immersion depth [54]. The intercept of the fitted line in Figure 9 indicates an immersion error of 0.05 mK when the probe head is at the bottom of the thermal well. A relative standard uncertainty of $0.18 \times 10^{-6}$ is assigned to account for immersion errors.

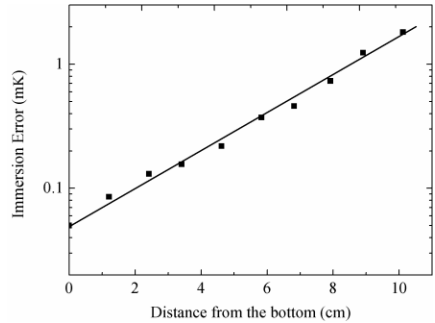

Figure 9: The immersion error of the resistor probe.

*4.4 Resistance measurement*
Equation (3) expresses Nyquist's law in terms of the resistance of the sensor, but strictly, it relates the noise power spectral density to the real part of the sensor impedance, $\mathrm{Re}(Z(f))$:

$$S_R = 4kT \, \mathrm{Re}(Z(f)). \tag{23}$$

Thus, in principle, the resistance or impedance of the sensor should be measured over the full operating bandwidth of the thermometer. While dc-resistance measurements can be made with relative uncertainties of 0.01 ppm or less [55], sensible measurements of resistance at the required accuracies and at ac, especially at frequencies above a few tens of kHz, are far more difficult [56]. For the JNT, the solution is to use a resistor with the widest practical bandwidth. This is achieved with very small thin film resistors manufactured with a meander pattern designed to balance the already small inductance and capacitance in such a way to maximise the bandwidth. The thin film construction also ensures that skin effects are minimised. This structure ensures that the ac frequency dependence of the noise spectrum is negligible and dominated instead by the inductance and capacitance of the connecting leads (transmission line), for which the frequency dependence is accommodated by the least-squares fit, as described in previous sections.

The dc resistance was measured with a DC resistance bridge and a standard resistor. The statistical relative uncertainty was typically $5\times10^{-8}$ for the average value of 200 individual measurements. The standard resistor is a hermetic encapsulated type with a calibration history traceable to the NIM quantum Hall resistance standard which maintained the value of SI ohm with an uncertainty of $2\times10^{-8}$. The drift rate of the standard resistor has been determined to be $3\times10^{-8}$/yr. The relative uncertainty in the standard resistance is estimated to be $1\times10^{-7}$ including its calibration and instability. The uncertainties from other effects, such as relaxation effect, thermoelectric, frequency dependence, had been analyzed in detail, and in this paper we assigned the same relative uncertainty to account for the effects [27].

4.5 Final result

With the analysis as described above, the fitted result of the noise-power ratio with a fourth-order fit over 550 kHz bandwidth, associated with the carefully calibrated value of the resistance $R$ and the temperature $T$ that are of traceability to the quantum Hall resistance and the current definition of the kelvin, and calculated value of the noise power $S_{Q\_calc}$, the value of the Boltzmann constant is determined to be $k = 1.3806514\times10^{-23}$ J/K, with a combined relative uncertainty of $3.5\times10^{-6}$. Table 1 summarises the contributions of all of various sources of uncertainty in the Boltzmann-constant determination. The dominate contribution is from the statistics and the spectral aberrations.

**Table 1:** Summary uncertainty budget for a determination of Boltzmann constant by QVNS noise thermometry. All uncertainties are expressed as relative uncertainties in parts per million.

| Component | Term | Relative uncertainty |
|---|---|---|
| Ratio of the power spectral densities, $S_R/S_Q$ | Statistical and spectral aberrations | 3.2 |
| | Dielectric losses | 1.0 |
| | Non-linearity | 0.1 |
| | EMI | 0.4 |
| | Total $u_r(S_R/S_Q)$ | 3.4 |
| TPW temperature $T$ | Reference standard TPW cell | 0.29 (0.08 mK) |
| | Temperature measurement | 0.04 (0.01 mK) |
| | Hydrostatic pressure correction | 0.08 (0.02 mK) |
| | Immersion effects | 0.18 (0.05 mK) |
| | Total $u_r(T_W)$ | 0.35 |
| Resistance $R$ | Ratio measurement | 0.05 |
| | Transfer Standard | 0.1 |
| | Ac-dc difference | 0.1 |
| | Relaxation effect | 0.5 |
| | Thermoelectric effect | 0.1 |
| | Total $u_r(R)$ | 0.53 |
| QVNS waveform $S_Q$ | Planck's constant | 0.044 |
| | Frequency reference | < 0.001 |
| | Quantisation effects | 0.1 |
| | Total($S_Q$) | 0.11 |
| | GRAND TOTAL ($k_B$) | 3.5 |

**5. Conclusion**

This paper reports the results of a measurement of Boltzmann constant, by Johnson noise thermometry, yielding a value of $k = 1.3806514(48) \times 10^{-23}$ J/K with a relative uncertainty of $3.5 \times 10^{-6}$. This result is $1.9 \times 10^{-6}$ high than the current (2010) CODATA value of Boltzmann constant, and within the uncertainty of the measurement. Although the uncertainty in this result is not as low as recent results obtained using acoustic gas thermometry [8-9], this measurement is purely electronic, based on different physical principles, uses different types of measuring equipment, and therefore provides very strong assurance that there are no major systematic errors affecting any of the recent $k$ determinations.

The measurement has been made possible by NISTs development of the quantum-accurate voltage noise source (QVNS) based on the ac Josephson voltage synthesiser. The QVNS enables (i) matching of the source impedances for the thermal and QVNS sources, which makes it possible to operate at wider bandwidths and reduced measurement times, (ii) a match of noise power enabling a reduction in the effects of non-linearities, (iii) a programmable noise source enabling linearity measurements, and (iv) a direct link to the fundamental constants underpinning the electrical units of measurement.

Several improvements have been made in the process of carrying out this measurement. Firstly, very close attention has been given to the matching of the frequency responses of the correlator to the thermal and QVNS signals, with the match achieved by including trimming resistors, capacitors and inductors to match the impedances of the transmission lines. Additionally, care was taken to minimise the dielectric loss in stray capacitances associated with the engineering of the thermal probe and the switch circuit.

Secondly, the entire JNT system is housed in an underground screened room remote from electromagnetic interference sources. For the first time, noise spectra have been obtained that are free of evidence of EMI. Subsidiary experiments using the QVNS at zero volts and a dummy thermal sensor confirm that any residual EMI effects are negligible.

Thirdly, greater attention has been given to the spectral analysis to achieve a better compromise between the complexity of the spectral model and the bandwidth in order to achieve the lowest practical uncertainty while eliminating the possibility of biases due to spectral mismatches. The analysis was facilitated by a new ADC board, operating with a Nyquist frequency of 2 MHz, which acquired more than 42 TB of data over a total integration time of 33 days.

Finally, the non-linearity of a noise thermometer has been measured directly for the first time. The non-linearity is remarkably low, and combined with a very close match of the correlated and uncorrelated noise powers, ensures the uncertainty due to non-linearity is negligible. The non-linearity measurements are made possible by the QVNS, which can provide quantum accurate noise voltages of (practically) any amplitude without modifying any other JNT operating condition.

We plan now to make a measurement of $k$ with a relative uncertainty below $3 \times 10^{-6}$ to meet the CCT second requirement (different physical technique) for the implementation of the new kelvin definition. This will be achieved by giving further attention to the spectral mismatches, and increasing the integration time.

ACKNOWLEDGMENT

The research team at NIM are deeply indebted to NIST for their support. The project was made possible with critical hardware, the QVNS chip fabricated by NIST, software, and the opportunity for team members to enjoy guest researcher positions at NIST, Boulder. The work at NIM is supported by NSFC (61372041 and 61001034) and the public welfare scientific research project (201010008).